# Why Everettians Should Appreciate the Transactional Interpretation

Ruth E. Kastner[†] and John G. Cramer[‡]

3/3/2010

ABSTRACT

The attractive feature of the Everett approach is its admirable spirit of approaching the quantum puzzle with a Zen-like "beginner's mind" in order to try to envision what the pure formalism might be saying about quantum reality, even if that journey leads to a strange place. It is argued that the transactional interpretation of quantum mechanics (TI), appropriately interpreted, shares the same motivation and achieves much more, with far fewer conceptual perplexities, by taking into account heretofore overlooked features of the quantum formalism itself (i.e. advanced states). In particular, TI does not need to talk about brain states, consciousness, or observers (rational or otherwise). In its possibilist variant ("PTI"), it shares the realist virtues of treating state vector branches as genuine dynamical entities, without having to explain how or why all of their associated outcomes actually happen (they don't), how to account for a plenitude of counterpart observers in some coherent notion of trans-temporal identity of the bifurcating observers (observers don't bifurcate in TI), nor how the certainty of all outcomes could be consistent with any coherent theory of probability, let alone the Born probability (the Born probability emerges naturally in TI). In short, TI is precisely the one-world interpretation Kent is looking for in his (2010).

## 1. Introduction and Background.

This paper will argue that the decades-long endeavor to explicate and fulfill Hugh Everett III's "Relative State Interpretation" (Everett 1957), while admirable in its motivation, determination and ingenuity, is nevertheless ill-fated, and that its original promise of a minimalist but faithful interpretation of quantum theory would be better fulfilled through a similarly open-minded and fervently dedicated exploration of Cramer's Transactional Interpretation (TI) (1980, 1986, 2005, 2006a,b; Kastner (2006, 2008, 2010).

[†] University of Maryland, College Park, Foundations of Physics Group
rkastner@umd.edu

[‡] Department of Physics, University of Washington, Seattle, WA 98195
jcramer@uw.edu

Kent (2010) has recently provided a carefully considered and somewhat gloomy assessment of the prospects for successfully fulfilling the Everettian program. While he expresses optimism "that we can find simpler one-world versions of quantum theory that have all the aforementioned virtues ['a well defined Lorentz covariant physical ontology that adds little or no arbitrary structure to the mathematics of quantum theory and that reproduces all the scientific successes of Copenhagen quantum theory within its domain of validity'] and none of the problems that afflict, and I think ultimately doom, the Everett program," he goes on to suggest (somewhat ominously, in our view) that "the failure of the Everett program adds to the likelihood that the fundamental problem is not our inability to interpret quantum theory correctly but rather a limitation of quantum theory itself." (2010, 2 (preprint version)) In this paper we argue against the latter conclusion.

We did not find in Kent's survey of the Everettian program any consideration of the issue of quantum nonlocality and its relation to the many-worlds view. Everett's original intent was to allow the formalism of quantum mechanics to lead him to a better and more rational way of dealing with the interpretational aspects of wave function collapse and measurement outcome.

In his original paper, Everett (1957) optimistically stated that "Fictitious paradoxes like that of Einstein, Podolsky, and Rosen, which are concerned with such correlated noninteracting systems, are easily investigated and clarified in the present scheme.". Despite that initial optimism, the approaches developed by Everett and his followers have usually halted with the consideration of a *single* measurement with many outcomes. The more general consideration of dual or multiple measurements on entangled quantum systems, as in EPR experiments, has not been adequately addressed. The interpretational problem with such systems is that while each individual measurement may have many possible outcomes, because of conservation laws acting in entangled systems, only certain outcomes of a second measurement are permitted for any particular outcome of the first measurement. Only pairs of outcomes are permitted for which the conserved quantity behind the entanglement (energy, momentum, angular momentum, spin projection, …) is properly conserved. It appears that Everett did not fully comprehend the central conundrum of nonlocality and entanglement.

The question raised by the many-worlds view is: for two measurements on entangled subsystems made at widely separated sites, how do the outcomes of the two measurements that are consistent with the entanglement conditions end up in the same "world"? One can envision two scenarios: (1) the first measurement to occur instantaneously splits off a world characteristic of a given outcome, and in that world the entanglements conditions restrict the possible outcomes of the second measurement to those consistent with the entanglement; or (2) each measurement creates a split of worlds, the splits somehow propagate (at light-speed?), meet in some intermediate region, and the splits of one measurement join or avoid the splits from the other measurement, linking up so that the entanglement conditions are respected. The problem with (1) is that it is inconsistent with special relativity. In most cases, the choice of which measurement occurs first depends on the reference frame from which the system is viewed. There is no unique "first measurement" that can create a world in which the second measurement can operate. So this approach manifestly violates Lorentz invariance. The problem with (2) is that propagating and self-sorting world splits and the mechanisms behind them are far removed from the formalism of quantum mechanics and from the spirit of Everett's minimalist approach to the interpretational problem. As far as we can tell, the Everettian program has produced no plausible account of how quantum nonlocality and multiple measurements on entangled systems should be viewed or interpreted. [1] We consider

---

[1] The prevailing Everettian approach thus far seems to be to deny nonlocality, arguing that EPR-Bell situations are "local" by denying counterfactual definiteness. This is legitimate as a starting point, but the account is then based on an obscure, Lorentz-noninvariant , and possibly subjectivist "splitting," as in

quantum nonlocality to be one of the central conundrums of quantum interpretation, and so we consider this to be a very serious failing.

The primary aims of this paper are (1) to follow up on Kent's optimism concerning a simpler one-world interpretation and (2) to dispel the notion that quantum theory might need *ad hoc* "fixing" because we are having trouble understanding its message about reality. Concerning (2), the situation is not that desperate: we *do* have the resources to make sense out of quantum theory in its well-corroborated pure form. We just need to reconsider a perfectly viable interpretation (TI) that has received insufficient attention, probably due to the particular kind of conceptual challenge it represents (i.e., time symmetry). Yet if current researchers are willing to countenance such admittedly 'fantastic' (Kent's term) or speculative features as bifurcating worlds and observers, 'probability' redefined as not requiring uncertainty about outcome [2], observer-dependent and ultimately subjective divisions of the world into 'system', 'observer', and 'environment',[3] the application of social philosophy and decision theory to subjectively defined 'rational' observers to try to derive purely physical laws such as Born probabilities, etc.[4], then surely they might be willing to reconsider a historically neglected interpretation (TI), especially if it makes much (if not all) of the foregoing unnecessary.

To review the basics of TI: an emitter emits an offer wave (OW), which formally corresponds to the usual quantum state vector. Depending on the experimental

---

Michael Clive Price's discussion at http://www.hedweb.com/everett/everett.htm#epr. Other approaches (e.g., as discussed in Rubin (2001)), involve using the Heisenberg picture and defining time-dependent operators that "carry" which-world information between quantum systems. This is marred by the conceptual problem of an endless list of proliferating labels which, again, is inconsistent with the minimalist Everettian approach and which seems to have no physical basis. We do not share Rubin's optimism that this problem will be adequately resolved by resorting to a quantum field theoretic account, since the basic phenomena are nonrelativistic (i.e., they are applicable to persistent particles at nonrelativistic speeds).

[2] As Peter Lewis (2007) notes, "Greaves (2004, 426–427) suggests giving up the assumption that a subjective probability measure [the weights appearing in the set of possible outcomes] over future events requires uncertainty about what will happen, and Wallace (2006, 672–673) suggests giving up the assumption that uncertainty requires some fact about which one is uncertain".

[3] As discussed in Schlosshauer (2007, 102)

[4] As discussed in extensive detail in Kent (2010).

arrangement, components of the OW (these would be analogous to "branches of the state vector in MWI) may be absorbed by one or more absorbers, each of which responds by sending an advanced (time-reversed) confirmation wave (CW) back to the emitter. (There is no counterpart of the CW in MWI or in any other prevailing interpretation, and that, it is suggested here, is the crucial missing ingredient in other attempts to interpret quantum theory.[5]) If there are $N$ such CW responses, there are $N$ incipient transactions in the form of OW/CW superpositions. The amplitude of the $i^{th}$ CW ($i = [1, N]$) at the locus of the emitter is equal to the Born probability, $|\langle\phi_i|\psi\rangle|^2$, of the outcome $\langle\phi_i|$ represented by the absorber in question conditional on the emitter state $|\psi\rangle$, which provides an immediate and unambiguous physical basis for the Born rule. So right away, we do away with any need for increasingly sophisticated, yet endlessly disputable, applications of decision theory and social philosophy to what may or may not be rational observers, in support of claims that such persons would *probably* arrive at the Born rule.

As noted above, the big conceptual barrier blocking many researchers from seriously considering TI is its time-symmetry: TI allows what Price (1997) terms "advanced action" in the form of confirmation waves; that is, physical influences (of limited scope) operating in a time-reversed direction. It is commonly thought that such influences must give rise to causal paradoxes, conflicts with relativity, or other inconsistencies. A specific example of such an assessment was Tim Maudlin's argument (1996, 2002) based on a "contingent absorber"-type thought experiment. In this experiment, one confirmation wave component for a slow-moving massive particle that is emitted in the form of offer waves (OW) in two possible directions (one to a near detector and one to a moveable farther detector behind it) would only be present if the nearby transaction failed, and the farther detector was then swung over to intercept the particle. Thus the farther detector's ability to return a CW was contingent on nondetection of the particle at the nearby detector. Maudlin then argued that the probability of detection based on this CW was only ½, yet the particle was certain to be detected there, which seemed inconsistent; and also, that the "pseudotime" account given by Cramer (1986) could not provide a coherent account of the process.

---

[5] The Bohmian interpretation does not suffer from the specific problem described here, but its account of the Born probability certainly falls short of the elegance and economy of TI's account.

However, Maudlin's objection has been answered by three different authors since then, all of them providing apparently reasonable ways for TI to remain viable in the face of this objection. Berkowitz (2002) argued that the Maudlin example constituted a causal loop, and argued that detection frequencies need not equal theoretical probabilities in the case of causal loops. Kastner (2006) argued that the Maudlin account argued against the original pseudotime account in Cramer (1986) but that this was merely a heuristic device and not fundamentally constitutive of TI; and that the core of TI was that the probability of an outcome was given by the weight of the incipient transactions, an approach that can be given a perfectly coherent account in a "big space" account of probability.[6] Cramer (2005) resurrected his "pseudotime" approach by proposing that all we need in such cases is a hierarchy in which possible transactions with shorter spacetime intervals have ontological priority over those with longer spacetime intervals, so that the nearby incipient transaction's outcome would have to be decided "before" (in pseudotime) other transactions could enter the competition. While one might not necessarily fully endorse any one of these proposals, it seems clear that, in view of three different ways to counter the Maudlin argument, his 1996 summary pronouncement of TI's "collapse" was at least premature.[7]

In a recent paper (Kastner 2010), one of the present authors argued that the best way to understand TI is in terms of a (modal) realist view of offer and confirmation waves[8]. This approach, termed therein "possibilist TI" (PTI), takes what Everettians would call "branches" of the state vector as representing real dynamical possibilities whose collective structure is described by Hilbert space (or Fock space in the relativistic domain). Thus the PTI already has much in common with the "Many Worlds" versions of Everett (MWI), which view branches of the state vector as real worlds in a "multiverse." The crucial differences between PTI and MWI are that in PTI, (1) absorbers play an equal role with emitters via time-reversed influences, as discussed above and (2) state vectors are viewed as physically real *possibilities* (but not actualities as in MWI).

---

[6] As defined in Placek and Butterfield (2002), Section 3.3.

[7] One of us (Kastner) is currently exploring yet another promising approach to address this type of "contingent absorber " objection based on dynamical properties of de Broglie waves.

[8] OW and CW are represented in PTI by pure state vectors (kets and bras respectively), rather than wave functions, the latter implying an *a priori* particular basis which is not appropriate in TI.

In the case of an entangled quantum system, this distinction between actuality and possibility is particularly significant. Much of the volume of such spaces describing the elements of an entangled quantum system is "illegal" because the components of the system described have variable values that are inconsistent because they violate the conservation laws acting in the system and producing the entanglement. Thus, most of the "branches" of the MWI multiverse are dead branches because conservation laws would not be respected within them. This is relevant to the point made above. The MWI is deficient in its treatment of entangled EPR-type systems. For the TI, only transactions that respect conservation laws are permitted to form. The MWI provides no similar Lorentz-invariant mechanism for pruning the dead branches.

The remainder of this paper will further explicate feature (2) and argue that PTI retains essentially all the virtues of Many Worlds Interpretations (MWI) while avoiding essentially all of the problems.

**2. The message of QM: possibility as a physically real resource.**

In support of an explicit many-worlds picture, David Deutsch has said:

"....quantum computers provide irresistible evidence that the Multiverse is real. One especially convincing argument is provided by quantum algorithms ... which calculate more intermediate results in the course of a single computation than there are atoms in the visible universe. When a quantum computer delivers the output of such a computation, we shall know that those intermediate results must have been computed somewhere, because they were needed to produce the right answer. So I issue this challenge to those who still cling to a single-universe worldview: if the universe we see around us is all there is, where are quantum computations performed? I have yet to receive a plausible reply." (1998)

Deutsch is right: the universe *is* more than what we see around us, but that does not mean that it has to be a multiverse, in which there are literally actual world counterparts to our own and in which all possible outcomes are actualized. The portion that we do not see, and that is responsible for the power of quantum computing over classical computing, can instead be interpreted as that which is *real but not actualized*: dynamical possibilities. That is, quantum computations can be processed through the medium of dynamical interacting possibilities without those processes having to be considered as actualized, observable outcomes. We *don't need* actualized intermediate outcomes in order to have physical room for these quantum computations, which can go on perfectly well (indeed, better)[9] "behind the scenes" to give rise to a final, actualized output. In the PTI, the output is what is detected (actualized), based on absorption of the post-computation OW and the ensuing CW, which provides for a transaction. The intermediate stages can be carried out by undetected offer waves (OW). So, all we have to do to reap the benefits of MWI without the problems is to consider branches of the state vector as real, yet *not actual* (the only actualized outcome being the final detected result).

This application of possibilist realism to quantum theory has ample (but overlooked) precedent in Heisenberg's comment:

> ``The probability wave of Bohr, Kramers, Slater...was a quantitative version of the old concept of "potentia" in Aristotelian philosophy. It introduced something standing in the middle between the idea of an event and the actual event, a strange kind of physical reality just in the middle between possibility and reality. (Heisenberg (2007), p. 15)

---

[9] That is, one could argue that it is the uncommitted (to a particular basis) nature of the offer wave which gives it its flexibility and thus its ability to explore "all possibilities at once."

Heisenberg never really pursued this bit of physical insight, but the PTI approach is to take his suggestion seriously: the state vector (an "offer wave" in TI) represents a kind of physical reality: that is, physically real possibilities "standing in the middle between the idea of an event and the actual event", which can interact with each other and with physical potentials and give rise to an actualized event by "setting the stage" for possible transactions, through which energy and other conserved quantities are transferred. It is these pre-detection interactions on the level of possibility that provide the extra information responsible for yet-to-be-realized power of quantum computing.

Here is another way to understand the power and efficacy of the "mere possibilities" represented by quantum states. Consider a hydrogen atom. The state vector of the electron can be seen as describing a distribution of positions (and a corresponding distribution of momenta) for the electron, but when no measurement is made on the electron—when it is not detected—its state can be considered a possibility wave "somewhere in the middle between possibility and actuality" in the sense described by Heisenberg above. In TI terms, it is an unabsorbed offer wave. Yet that "mere possibility" is incredibly powerful — powerful enough to support the structure of matter and to provide its apparent solidity. Note that in his (1998), Deutsch wants to describe such an electron as existing in all his many (interfering) worlds—being actualized in all possible different outcomes in separate worlds. But since no observational basis has been specified, are these many worlds ones in which the electron has a definite (more precisely, narrowly localized) position, i.e. a splitting with respect to the position basis? Or momentum? [10] *How much simpler it is to just view the state vector as representing a real and potent (if not 'actual') entity sufficient in itself, uncommitted to any particular basis.* Again, the point is that we *don't need* to posit actualized worlds corresponding to specific outcomes (and then have to worry about the ambiguity of basis for these outcomes) to get the job done, if we view possibilities—represented by state vectors--as having dynamic potency.

---

[10] Of course, position and momentum are Fourier transforms of each other; eigenfunctions of the "position basis" are commonly understood to be represented by Dirac delta functions. In either case you need an infinite number of worlds of one kind to correspond to an eigenstate of the complementary observable.

Admittedly, there is "collapse" in the TI (or PTI). However, note that the collapse is completely "defanged" when compared to the usual notion of collapse. First, there is no need for an observer: collapse occurs any time an emitter receives one or more CW in response to an OW. The new interpretational ingredient that "cuts the Gordian knot" of the apparent observer-dependence of quantum phenomena is the taking into account of the dynamical role of absorbers on an equal footing with emitters. Thus, the TI is an "observer-free collapse interpretation", in Bub's terms (1997). In TI, collapse is not observer-dependent but simply *absorber*-dependent. Collapse is the formation of a transaction.

Without including the role of the absorber, all we have is an offer wave (the "quantum state") that never gets a response, so it is typically considered to be propagating endlessly out into the world, continually being amplified depending on what it happens to encounter: a Geiger counter; a cat; an observer; Wigner; Wigner's Friend; Alice; Bob; etc; etc;. *ad infinitum*. Without taking absorbers into account, there is no principled way to call a halt to this proliferating, ever-amplifying quantum state. In the farthest extreme, we have the "universal state vector" unitarily propagating ever onward, so that (from a God's eye view), allegedly nothing ever "really" happens. Thus arises the necessity to consider arbitrary divisions of the universal state vector into "observer" and "observed." All of this is avoided in TI: absorbers provide the confirmation waves that give rise to incipient transactions (four-vector superpositions of OW and CW) and thus trigger collapses (actualized transactions), bringing clear and decisive closure to state vectors at appropriate levels. Such actualized transactions will furthermore naturally line up with decoherence arguments, since decoherence, taking into account as it does the total environment of a system, is fundamentally based on the nature of absorbers available to emitted particles.

Deutsch, as quoted above, challenges advocates of a single-universe worldview to explain where quantum computations are performed and more specifically, how they could possibly be performed in a single universe. The answer is, they are performed *right here* (at the level of possibility, as noted earlier) From the PTI viewpoint, a quantum computer is simply a quantum system that has been set up with quantum-logical circuit constraints such that the only transaction that is allow to form is the one that produces the desired calculational result by satisfying all of the constraints. The offer waves generated at the input propagates through the system and "feels it out" in parallel along all of its branches and circuit elements. Some path through the maze of quantum logic produces a consistent OW/CW echo back at the input. Then the appropriate transaction (which has no competitors because of the constraints) forms, and the quantum computer delivers the desired answer. We find this one-world picture of quantum computation to be much more satisfying than the Deutsch scenario of somehow doing parallel calculations in a vast array of ephemeral branch universes.

## 3. What is the "pure" theory and why does TI address it effectively?

In Everett's view and that of his followers, the "pure theory" is only the unitary evolution of the state vector, without the projection postulate. But an important part of the theory—the part that allows it to make empirical contact with experience—is the Born Rule. Thus the "pure theory" should properly be considered to be the *combination* of linear evolution of the state vector with the Born Rule. The Born Rule cannot just be tacked on as an afterthought: it is a crucial component of the theory, just as a crucial part of electromagnetic theory is that the electrical (or magnetic) energy of the field is proportional to the square of the field. Nobody would try to interpret electromagnetic theory by initially ignoring the expression for electrical energy just because (hypothetically) it was not clear how that quantity was physically related to the field, and then trying to account for the energy after the fact by considering "FAPP"-type explanations such as what a rational observer might expect to bet on when making decisions about electric field-based phenomena. The genuine interpretational challenge would be to understand how the electrical energy is physically related to the field. The same challenge applies to quantum theory: how is the probability for outcomes *physically* related to the state vector? Everettian approaches can give no answer in these terms, as Kent's discussion makes clear.

So the pure theory properly consists of *both* the linear evolution of the state vector and the well-corroborated empirical link with experience, the Born Rule. We need a *physical* explanation for the Born Rule. TI provides a simple and elegant one: the Born Rule corresponds to the final amplitude of the CW at the locus of the emitter. Since it is the transaction based on that CW component that may, or may not, result in actualization of the corresponding outcome, we have a genuinely probabilistic situation: an objectively uncertain result whose probability is precisely the amplitude of the CW at the emitter.

## 4. Conclusion: TI deserves serious and open-minded reconsideration.

It has been argued that there is in fact a perfectly viable one-world (one *actual* world, that is) interpretation that can potentially fulfill Kent's requirements for "a mathematically elegant, universally applicable, Lorentz invariant, scientifically adequate [interpretation] of quantum theory that supplies a well-defined realist ontology." (Kent 2010, 2). As noted in the Abstract, TI approaches the *total* formalism--the combination of the linear unitary evolution and the Born Rule--with a "Beginner's Mind" and tries to see what the formalism is really telling us. Clearly, the mathematics expresses a symmetry of retarded and advanced solutions, and truly "letting the formalism speak" means not assuming that part of it (the advanced part) should be ignored because it doesn't fit our metaphysical preconceptions about the directionality of time. Taking into account the full content of the formalism, TI provides the following: a straightforward, simple and elegant account of the Born Rule; an observer-free account of collapse; the collapse is Lorentz-invariant since it occurs either atemporally or all along a spatiotemporal four-vector (depending on one's ontological interpretation); a realist ontology in terms of possibilist realism, thus providing a clear answer to "where all the computation takes place" in quantum computing. It potentially opens a door to an entirely new and exciting understanding of physical reality worthy of the great empirical successes of quantum theory: namely, that the world around us is seething with unactualized, but nevertheless real and potent, physical possibility.

While no pretense has been made here that the ontology of TI (in Cramer's account) or PTI (in Kastner's account) is completely worked out, it seems clear that there is much fruitful ground to explore. We hope that some of the great minds currently engaged in what (to us, and apparently to others as well, e.g. Kent) appears to be a losing battle with the Everett approach will consider devoting themselves instead to exploring the exciting possibilities inherent in the Transactional Interpretation.